\documentclass[11pt,aps,amsmath,amssymb,nofootinbib,notitlepage,longbibliography]{revtex4-1}
\usepackage{amsmath,amssymb}
\usepackage{slashed}
\usepackage{graphicx}
\usepackage{bm}
\usepackage[top=1.0in,bottom=1.0in,left=1.0in,right=1.0in]{geometry}
\usepackage[colorlinks,linkcolor=blue,citecolor=blue]{hyperref}

\newcommand{\be}{\begin{equation}}
\newcommand{\ee}{\end{equation}}
\newcommand{\bea}{\begin{eqnarray}}
\newcommand{\eea}{\end{eqnarray}}
\newcommand{\ba}{\begin{eqnarray}}
\newcommand{\ea}{\end{eqnarray}}

\begin{document}

\title{Mass sum rule of hadrons in the QCD instanton vacuum}


\author{ Ismail Zahed}
\email{ismail.zahed@stonybrook.edu}
\affiliation{Center for Nuclear Theory, Department of Physics and Astronomy, Stony Brook University, Stony Brook, New York 11794--3800, USA}



\begin{abstract}
We briefly review the key aspect of the QCD instanton vacuum in relation to the quantum breaking of conformal symmetry
and the trace anomaly.  We use  Ji$^\prime s$  invariant mass decomposition of the energy momentum tensor together with the trace anomaly,
to discuss the mass budget of the nucleon and pion in the QCD instanton vacuum.  A  measure of the gluon 
condensate  in the nucleon, is a measure of the compressibility of the QCD instanton vacuum as a dilute
topological liquid.
\end{abstract}

\maketitle

\section{Introduction}
A remarkable feature of QCD is that in the chiral limit it is a scale free theory. Yet,  all hadrons are massive,  composing
most of the visible mass in the universe. The typical hadronic scale is 1 fm, but where does it come from? The answer appears to be  from
a subtle quantum effect referred to as dimensional transmutation, and  related to the quantum breaking of the conformal symmetry of QCD.
This mechanism is non-perturbative. On the lattice, the lattice cutoff along with the running coupling 
combine to generate this scale. In the continuum, to achieve this mechanism requires a non-perturbative description of the
vacuum state and its excitations. 

The QCD vacuum as a topological  liquid of instantons and anti-instantons,  offers by far the most compelling non-perturbative
description that is analytically tractable in the continuum, thanks to its QCD semi-classical  origin  and 
diluteness~\cite{Diakonov:1995ea,Schafer:1996wv,Nowak:1996aj}. It is not the only description. 
Other candidates based on center vortices and  monopoles to cite a few~\cite{Greensite:2016pfc}, 
are also suggested and may as well be present in addition to the instantons. However, the latters appear to trigger
the dual  breaking  of conformal and chiral symmetry breaking, and dominate the vacuum state and its low-lying
hadronic excitations. Center vortices maybe important for the disordering of the large Wilson loops and confinement, 
a mechanism likely at work in the orbitally excited hadrons as they Reggeize.

The spontaneous breaking of chiral symmetry rather than confinement drives the formation of the low-lying and 
stable hadrons such as the nucleon and pion. In the QCD instanton vacuum conformal symmetry is broken by the
density of instantons: {their continuous  quantum rate of tunneling in the vacuum}. The  breaking of chiral symmetry follows simultaneously  from the
delocalization of the light quark zero modes, by leapfrogging the instantons and anti-instantons much like electrons
leapfrogging atoms in a metal. Detailed numerical simulations of light hadronic correlators in the QCD instanton
vacuum~\cite{Shuryak:1999fe} show remarkable agreement with direct lattice measurements~\cite{Chu:1993cn},
and a wealth of correlators extracted from data~\cite{Shuryak:1999fe}. The universal conductance fluctuations in the zero mode
region of the Dirac spectrum, 
predicted by random matrix theory~\cite{Verbaarschot:1993pm} and confirmed by lattice simulations~\cite{Wittig:2020jtm}, 
show  unequivocally the topological  character of the origin of mass.

In this note, we briefly review the salient aspects of the QCD instanton vacuum in relation to the quantum breaking of
conformal symmetry in section~\ref{sec_instantons}. We then discuss the role of the trace anomaly  in the  nucleon and
pion mass in section~\ref{ANOMALY}. The quark and gluon composition of the hadronic mass using Ji$^\prime s$
decomposition~\cite{Ji:1994av} is discussed in section~\ref{sec_hamiltonian}. In section~\ref{sec_compressibility}, 
we show that the gluon condensate in the nucleon is  tied to the QCD vacuum compressibility, a measure of the
diluteness of the QCD instanton vacuum as a topological liquid. Our conclusions are  in section~\ref{sec_conclusions}.

\section{QCD instanton vacuum}
\label{sec_instantons}

\begin{figure}[h!]
	\begin{center}
		\includegraphics[width=16cm]{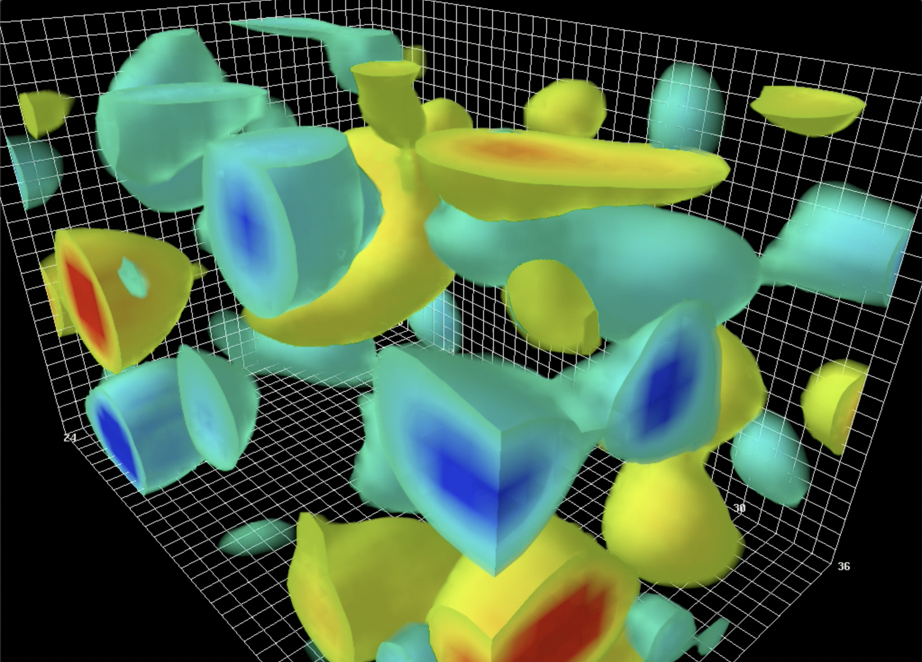}
		\caption{Instantons (yellow) and anti-instantons (blue) configurations in the cooled Yang-Mills  vacuum~\cite{Leinweber:1999cw}.
		They constitute the primordial gluon epoxy at the origin of  the hadronic mass. See text.}
		\label{fig_VAC}
	\end{center}
\end{figure}

As we noted above, the chief aspect of the QCD vacuum (meaning quenched throughout)  is its quantum breaking of conformal symmetry with the emergence of all light
hadronic scales. The nature of the gauge fields at the origin of this breaking were mysterious and the subject of 
considerable debates and speculations for many decades, till  stunning pictures were developed by Leinweber 
and his collaborators using cooling and/or projection techniques~\cite{Leinweber:1999cw,Biddle:2019gke}. Out of the fog of millions of gauge fluctuations, 
cooling has revealed a stunning vacuum landscape composed of inhomogeneous and topologically active gauge fields as shown in
Fig.~\ref{fig_VAC}. Remarkably, the key features of this vacuum were predicted long ago by Shuryak~\cite{Shuryak:1981ff}

\begin{equation}
n_{I+\bar I}\equiv \frac 1{R^4}\approx \frac 1{ {\rm fm}^{4}} \qquad\qquad\frac{\bar \rho}R \approx  \frac 13   \label{eqn_ILM}
\end{equation}
for the instanton plus anti-instanton density and size, respectively. In other words, the hadronic scale $R=1\,{\rm fm}$
emerges as  the mean quantum tunneling rate of the topological charge in the QCD vacuum.
The dimensionfull parameters (\ref{eqn_ILM})
combine in the  dimensionless packing parameter
$\kappa\equiv \pi^2\bar\rho^4 n_{I+\bar I}\approx 0.1$,
a measure of  the diluteness of the instanton-anti-instanton ensemble in the QCD vacuum. 
Fortunatly, the smallness of $\kappa$ is what will allow us to do reliable analytical calculations. 
In the cooled landscape shown in Fig.~\ref{fig_VAC}, most hadronic correlations are left unchanged with those 
before the cooling takes place
~\cite{Schafer:1996wv} (and references therein).

 The size distribution of the instantons
and anti-instantons density (their tunneling rate per   size) in the QCD vacuum is well captured semi-empirically by

\begin{equation}
\label{dn_dist}
\frac{dn(\rho)}{d\rho} \sim  {1 \over \rho^5}\big(\rho \Lambda_{QCD} \big)^{b} \, e^{-\#\rho^2/R^2}
\end{equation}
with $b=11N_c/3-2N_f/3$ (one loop).  The small size distribution follows from the conformal nature of the
instanton moduli and perturbation theory. The large size distribution is non-perturbative. A variational analysis of the
QCD instanton vacuum including binary interactions~\cite{Diakonov:1995ea}, shows that the large size instantons are cutoff self-consistently 
by their density as in (\ref{dn_dist}). Lattice parametrization of the same distribution suggests that the cutoff is due to
the onset of confinement with $R\approx 1\,{\rm fm}\rightarrow l_s\approx 0.2\,{\rm fm}$ (the string length)~\cite{Hasenfratz:1999ng,Shuryak:1999fe}.  
Here, we favor the former cutoff as it preserves the strictures of the renormalization group invariance following
the quantum  breaking of conformal symmetry, with a single scale overall in the chiral limit ($R$ sets the scale for both the gluon and chiral condensates).

\subsection{Quantum conformal symmetry breaking and the trace anomaly}

 The quantum breaking of conformal symmetry is best captured through the trace of the energy
momentum tensor. Indeed,  consider its symmetric form

\bea
\label{1}
T^{\mu\nu}=&&\frac 2{\sqrt{-g}}\frac{\delta S_{1+3}}{\delta{g_{\mu\nu}}}
=F^{a\mu\lambda}F^{a\nu}_\lambda-\frac 14 g^{\mu\nu}F^2+\frac 14 \overline\psi \gamma^{[\mu} i\overleftrightarrow D^{\nu]_+}\psi
\eea
with $\overleftrightarrow{D}=\overrightarrow{D}-\overleftarrow{D}$ and $[]_+$ denotes symmetrization. 
It is conserved $\partial_\mu T^{\mu\nu}=0$, with an anomalous  trace

\be
\label{3X}
T^\mu_\mu=\frac{\beta(g^2)}{4g^4}F^a_{\mu\nu}F^{a\mu\nu}+m\overline\psi\psi
\ee
with  the Gell-Mann-Low beta-function (2 loops)

\be
\label{beta}
\beta(g^2)=-\frac{bg^4}{8\pi^2}-\frac{\bar bg^6}{2(8\pi^2)^2}+{\cal O}(g^8)
\ee
Throughout, we use the rescaling $gF\rightarrow F$ for all operators in the instanton and anti-instanton gauge fields.
In the QCD instanton vacuum, the gluon operator
 $F^2/(32\pi^2)\rightarrow (N_++N_-)/V$ counts the number of instantons plus anti-instantons
per 4-volume $V$. In the canonical ensemble  with zero theta-angle, it is fixed by the instanton density  with $N_\pm/V=\bar N/2V$. 
Therefore  we have

\be
\label{SCALE}
\left<T^\mu_\mu\right>\approx -b \bigg(\frac{\bar N}{V}\bigg)+m\left<\overline \psi\psi\right>\approx
-b \bigg(\frac{\bar N}{V}\bigg)\bigg(1+{\cal O}(mR)\bigg)\approx -10\,{\rm fm}^{-4}
\ee
setting the scale of all hadrons. 
The current mass $m\approx 8$ MeV in (\ref{SCALE})  is fixed  at the soft renormalization point 
$\bar \rho\approx 0.3$ fm, about twice  the  commonly used value at the hard renormalization scale.
The scale of the spontaneous breaking of chiral symmetry is also  fixed by  the finite instanton density, but its contribution
to the vacuum scalar density  is small  since  $mR \approx (8\,{\rm MeV})(1\,{\rm fm})\approx 1/25$.

Since the vacuum is Lorentz symmetric, (\ref{SCALE}) amounts to a
negative vacuum energy density $B=\left<T^\mu_\mu\right>/4\approx -(250\,{\rm MeV})^4$,
with no strict confinement  at work.
Note that $B\approx -b\left<F^2\right>$, where the important overall negative sign inherited from the scale anomaly
 is ultimatly a quantum magnetic
effect (sign of the beta function). The gluon condensate $\left<F^2\right>$ is always positive in Euclidean
signature. This is usually referred to as the gluon epoxy  (a term coined by the late Gerry Brown).

Some of the quantum  scale fluctuations in QCD are captured in the QCD instanton vacuum
using the grand-canonical description instead of the canical one. In the former, the  instanton number $N=N_++N_-$ is allowed to fluctuate with the measure
\cite{Diakonov:1995qy,Kacir:1996qn,Nowak:1996aj}

\be
\label{dist}
\mathbb P(N)=e^{\frac{bN}4 }\bigg(\frac {\overline N}{N}\bigg)^{\frac {bN}4 }
\ee
which is stronger than Poisson ($b/4\rightarrow 1$),  to reproduce the vacuum compressibility

\be
\frac{\langle(N-\bar N)^2\rangle_{\mathbb P}}{\bar N}=\frac 4b 
\ee
expected from QCD low-energy theorems~\cite{Novikov:1981xi}.

\subsection{Spontaneous breaking of chiral symmetry and conductance fluctuations}

The  quantum  breaking of conformal symmetry and
the generation of the $R=1$ fm and a gluon condensate,  is a direct measure of the instanton tunneling rate or topological density 
in the  vacuum. In the quenched approximation, it is solely a property of the gluon fields.
This is a necessary but not a sufficient mechanism for  hadronic mass generation. The sufficient mechanism, 
which relies on these topological  fields, is the delocalization of the quark zero modes and the ensuing 
spontaneous breaking of chiral symmetry. The result is a vacuum chiral condensate and the emergence  of a quark mass,
both of which are fixed by the same $R=1$ fm scale at the origin of the gluon condensate. 
This topological mechanism for mass generation 
leaves behind  a fingerprint: universal conductance-like fluctuations in the  quark spectrum~\cite{Verbaarschot:1993pm}.
This is a lesser known fundamental phenomenon,  that we now briefly discuss.

\begin{figure}[h!]
	\begin{center}
		\includegraphics[width=10cm]{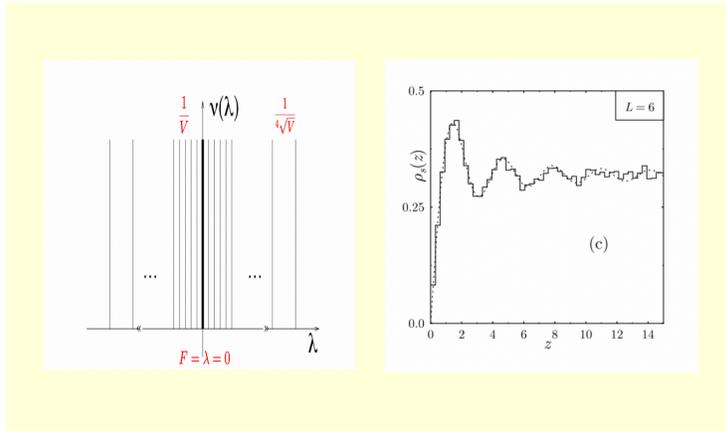}
		\caption{Left: sketch of the virtual Dirac spectrum for the disordered quark zero modes. Right:  the universal spectral or conductance fluctuations (dotted line)  predicted by 
		random matrix theory (\ref{NUS})~\cite{Verbaarschot:1993pm}, 
		and measured in lattice generated gauge configurations (solid line)~\cite{Wittig:2020jtm}. See text.}
		\label{fig_ZERO}
	\end{center}
\end{figure}

 Decades ago, Banks and Casher observed
 that the spontaneous breaking of chiral symmetry with a finite chiral condensate $\langle \overline\psi\psi\rangle$,
  is  associated to a huge accumulation of the quark zero modes near the zero point of the virtual quark spectrum
as  illustrated    in Fig.~\ref{fig_ZERO} (left), and  captured  by the relation~\cite{Banks:1979yr}
 
 \bea
 \label{BK1}
 \langle \overline  \psi  \psi\rangle =-\pi \nu(0)\equiv -\sigma_C
 \eea
The quark density of states is

\bea
\label{NUA}
\nu(\lambda)=\lim_{m\to 0}\lim_{V\to \infty}\frac 1{V}\left<\sum_n\delta(\lambda-\lambda_n[A])\right>_A\equiv \frac 1{V}\frac 1{\Delta\lambda}
\eea
with the virtual quark eigenstates solution to 
$$(iD[A]+im)q_n[A]=(\lambda_n[A]+im)q_n[A]$$ for a given gauge configuration.
A finite $\nu(0)$ means that $\Delta\lambda\approx R^3/V$, as opposed to $\Delta\lambda\approx 1/{}^4\sqrt V$ in the continuum. The quark spectrum is extremely
dense near $\lambda=0$, as the disordering turn the quark zero modes to quasi-zero modes.

Remarkably, the Banks-Casher relation (\ref{BK1}) resembles the Kubo formula for 
the DC conductivity  in metals  with $\sigma_C\leftrightarrow  -\langle \overline\psi\psi\rangle$ and  the zero-virtuality point $F=\lambda=0$ playing the role of the Fermi-surface.
The QCD vacuum turns metallic when chiral symmetry is spontaneously broken. The same holds for the QCD instanton vacuum. In the unquenched case, this statement
is extremely non-trivial. A  too dilute or too dense topologically active vacuum would result into either a neutral gas of instanton-anti-instanton molecules, or a crystal arrangement
of instantons-anti-instantons,  
instead of a liquid, with no spontaneous
breaking of chiral symmetry~\cite{Diakonov:1995ea,Schafer:1996wv,Nowak:1996aj}. We are thankfull that nature has served us a dilute liquid!

In fact there are infinitly many Banks-Casher-like formula which capture the fluctuations of the conductance $\sigma_C$ in the mesoscopic
limit, with the connected  moments $\sigma_C^n= \langle (\overline\psi\psi)^n\rangle_C $.  These conductance fluctuations are universal, and follow from 
chiral random matrix theory, with the result for the mesoscopic spectral density at zero virtuality given by the master formula~\cite{Verbaarschot:1993pm}

\bea
\label{NUS}
\nu_s(z=N\lambda)=\frac{z}2\bigg(J^2_{N_f}(z)-J_{N_f+1}(z)J_{N_f-1}(z)\bigg)
\eea
with $N, V\rightarrow \infty$ but fixed $z=N\lambda$ and $N/V$.
These predicted conductance fluctuations were later  measured in numerically  generated lattice gauge configurations as  shown in Fig.~\ref{fig_ZERO}~\cite{Wittig:2020jtm} (right). 
The dotted line is the result (\ref{NUS}) and the solid line is the lattice measurement.
This is  one of the signature for the topological origin of the hadronic mass scale, as it centers on the quark zero mode zone. The zero modes only occur in the 
presence of topological gauge fields such as the instantons and  anti-instantons~\cite{tHooft:1976snw} (or their close cousins the instanton-dyons~\cite{Chernodub:1999wg,Liu:2015jsa}), 
a consequence of the Atiyah-Singer index theorem (or the Atiyah-Patodi-Singer for their cousins).

\section{Mass  identity}~\label{ANOMALY}

We now focus on the anomalous trace of the QCD energy momentum tensor,  and its relation to the 
hadron mass. It is worth stressing that the ensuing relation to the mass is just a bulk identity and not a mass decomposition.
Having said that, the trace couples to a scalar dilaton which sources a sigma (2-pion) meson and/or a $0^{++}$ scalar glueball field.
QCD perturbative arguments suggest that this trace may be  accessible in the photo-production of charmonium 
at  treshold~\cite{Kharzeev:2021qkd,Joosten:2018gyo} (and references therein), although the coupling to the $2^{++}$  tensor glueball may still be  
very active in the treshold region~\cite{Mamo:2019mka}.  Recall that the Reggeized form of the $2^{++}$ exchange transmutes to the Pomeron,
 and is dominant asymptotically.

\subsection{Nucleon}

For a  nucleon  state $|P\rangle$ with the standard normalization $\langle P|P^\prime\rangle=2E_P(2\pi)^3\delta^3(P-P^\prime)$, one has

\begin{align}\label{forwardT}
\langle P|T^{\mu\nu}|P\rangle=2P^{\mu}P^{\nu} \ .
\end{align}
with the  trace in any frame (one-loop)

\begin{align}\label{Trace}
\langle P|T^\mu_\mu|P\rangle=\langle P|\bigg(-\frac {b}{32\pi^2} F^2+m\bar \psi\psi\bigg)|P\rangle=2M_N^2 \ .
\end{align}
with  $g^2 F^2\rightarrow F^2$  for the strong instanton and anti-instanton gauge fields. It is renormalization group invariant. 
The identity (\ref{Trace}) shows that the nucleon mass is the change of the conformal anomaly or gluon field in a nucleon state.
However, the formation of the state occurs only  if chiral symmetry is spontaneously broken as we noted earlier.
 (\ref{Trace})  is a  QCD identity that is satisfied in the QCD instanton 
vacuum as we now show.

In the rest frame, the gluon contribution in (\ref{Trace}) 
follows from the normalized and connected 3-point function asymptotically

\be
\label{corr}
\frac{\langle P|F^2|P\rangle}{\langle P|P\rangle}={\lim_{T\to\infty}}\frac{\left<J_P^\dagger (T)F^2J_P(-T)\right>_C}{\left<J_P^\dagger (T)J_P(-T)\right>}
\ee
with $J_P$ a pertrinent nucleon source. In the canonical description 
of the QCD instanton vacuum $F^2/(32\pi^2)\rightarrow \bar N/V$ is a  number. 
It factors out in the 3-point correlator in  (\ref{corr}) (numerator), and the connected correlator vanishes. 

A non-vanishing contribution to the connected 3-point correlator follows from the 
grand-canonical  description,  where $N$ is  allowed to fluctuate  as we noted in (\ref{dist}).
With this in mind, it is straightforward to see that (\ref{corr}) is dominated by the variance

\bea
\label{corr1}
\frac V{32\pi^2}\frac{\langle P|F^2|P\rangle}{\langle P|P\rangle}\approx \left<(N-\bar N)^2\right>_{\mathbb P}\frac{\partial}{\partial \bar N}{\rm Log}\bigg({\lim_{T\to\infty}}{\left<J_P^\dagger (T)J_P(-T)\right>}\bigg)
\eea
with the higher moments suppressed by $1/b^2\sim 1/N_c^2$. The result (\ref{corr1}) was noted in~\cite{Diakonov:1995qy}  (see Eq. 5.8) 
using a fermionization method, 
and in~\cite{Kacir:1996qn} (see Eqs. 91,93) using a bosonization method, each  of the QCD instanton vacuum in the $1/N_c$ approximation. 
The expectation value in the first bracket is carried using the distribution (\ref{dist}). (\ref{corr1}) illustrates how the nucleon
scoops the epoxy from the QCD instanton vacuum.

All dimensions in the QCD instanton vacuum are fixed
by the density $\bar N/V=1/R^4$ and the current quark masses. 
The nucleon mass is the sum of the chirally symmetric  (invariant mass) plus the symmetry breaking contribution (pion-nucleon sigma term),

\be
\label{MASS}
M_N=M_{\rm inv}+\sigma_{\pi N}=  C\bigg(\frac {\bar N}V\bigg)^{\frac 14} +\bar C m \bigg(1+{\cal O}(mR)\bigg)
\ee
with~\cite{Steele:1995yr,Alexandrou:2020okk,Hoferichter:2016ocj}
 
 \be
\label{PIN}
\sigma_{\pi N}=\frac{\left<P|m\overline\psi\psi|P\right>}{\left<P|P\right>}\approx 50\,{\rm  MeV}
\ee
evaluated at the soft renormalization scale $\bar\rho=0.6\,{\rm GeV}$, which is  the appropriate scale for hadronic spectroscopy.
The right-most relation in (\ref{MASS}) follows from the QCD instanton vacuum. As a result,
 the anomalous contribution in the QCD instanton vacuum  is

\bea
\label{corr2}
\frac V{2T}\frac {-b}{32\pi^2}\frac{\langle P|F^2|P\rangle}{\langle P|P\rangle}
=4\frac{\partial M_N}{\partial {\rm Log}\bar N}=M_{\rm inv}
\eea
which is seen to satisfy the sum rule 

\bea
\frac{\left<P|T^\mu_\mu|P\right>}{2M_N}=M_{\rm inv}+\frac{\left<P|m\overline \psi\psi|P\right>}{2M_N}=M_N
\eea


\subsection{Pion}

The preceding arguments apply also to the pion, with one major difference, 

\bea
m_\pi=C\sqrt{m}\bigg(\frac{\bar N}{V}\bigg)^{\frac 18}(1+{\cal O}(mR))
\eea
since it is a Goldstone mode. 
The ${\cal O}(mR)$ corrections are small  in the QCD instanton vacuum. 
It follows that 

\bea
\label{corr2}
\frac V{2T}\frac {-b}{32\pi^2}\frac{\langle \pi|F^2|\pi\rangle}{\langle \pi|\pi\rangle}
=4\frac{\partial m_\pi}{\partial {\rm Log}\bar N}=\frac 12 m_\pi
\eea
which was first observed in~\cite{Ji:1995sv}, 
with the sum rule

\bea
\frac{\left<\pi|T^\mu_\mu|\pi\right>}{2m_\pi}=\frac 12 m_\pi+\frac{\left<\pi|m\overline \psi\psi|\pi\right>}{2m_\pi}=m_\pi
\eea
satisfied, as expected.
The pion sigma-term follows from chiral reduction or the Feynman-Hellmann theorem

\be
\frac{\left<\pi|m\overline \psi\psi|\pi\right>}{2m_\pi}=\frac{\partial {\cal E_\pi}}{\partial {\rm log}m}=\frac 12 m_\pi
\ee

\section{Ji  mass sum rule}~\label{sec_hamiltonian}

The trace identity (\ref{forwardT}) reflects on the general fact that all hadron masses in QCD  are tied to the quantum  breaking of
conformal symmetry as we noted earlier, and should be enforced by any non-perturbative quantum description, wether numerical such as the lattice
or analytical such as the QCD instanton vacuum. 
However,  it does not specifically budget this mass breaking in terms of  the hadron  constituents. 
In a strongly interacting  theory,  this issue may be elusive, especially with a soft renormalization scale,
as the gluons are strongly untertwined with the light quarks. This is more so
in the unquenched and screened formulation.

This notwithstanding, a specific and physically  motivated proposal to budget the mass, was 
put forth by Ji in~\cite{Ji:1994av,Ji:1995sv},  and  since revisited by many~\cite{Lorce:2017xzd,Roberts:2021xnz,Metz:2020vxd} (and references therein). The ensuing mass composition
involves the sum of partonic contributions, some of which may be measurable  using DIS experiments.
The proposal relies on an invariant decomposition of the energy momentum tensor which we now detail.

The  energy-momentum tensor (\ref{1}) can be decomposed as the sum of a traceless and tracefull part~\cite{Ji:1995sv,Ji:2021}

\begin{align}
\label{T1}
T^{\mu\nu}\equiv \bar T^{\mu\nu}+\hat T^{\mu\nu}\equiv \bar T^{\mu\nu}+ {g^{\mu\nu}}\,\frac  14{T}^\alpha_\alpha \ ,
\end{align}
where the traceless part reads

\begin{align}
\label{T2}
\bar  T^{\mu\nu}=\bigg(-F^{a\mu\tau}F^{a\nu}_\tau+\frac 14 g^{\mu\nu}F^2\bigg)
+\frac 14 \overline\psi \gamma^{[\mu} i\overleftrightarrow D^{\nu]_+}\psi-{g^{\mu\nu}}\frac 1{4}m\bar\psi \psi ,
\end{align}
and the tracefull  part is 

\be
\label{T3}
{\hat T}^{\mu\nu}=g^{\mu\nu}\frac 14\bigg(\frac{\beta(g^2)}{4g^4}F^2+m\overline\psi\psi\approx 
-\frac {b}{32\pi^2}F^2+m\overline\psi\psi\bigg)
\ee
 We note that this decomposition is commensurate with the analysis of the nucleon energy momentum tensor in
 holographic QCD, through dual gravitons in bulk~\cite{Mamo:2019mka}. (Holography is  a good example of a strong
 coupling description of a gauge theory via its  gravity dual,   where the partonic structure is elusive).

The tracefull and traceless part of the
energy momentum tensor  (\ref{T1}-\ref{T3}) correspond to the spin-2 and spin-0 representations of the Lorentz group, and do not mix 
under renormalization by symmetry. 
Their renormalization  at  the  instanton size scale $\bar\rho\approx 0.3\,{\rm fm}$ is subsumed  throughout. On  the
lattice, this soft renormalization scale is best achieved using 
a cooling procedure where only the  UV quantum and non-singular fluctuations are subtracted (our instantons are classical  fields in 
singular gauge!).  Note that our renormalization scale is softer than the one  used in currently  fine lattices with $1/\mu\approx 0.1\,{\rm fm}$
($\overline{\rm MS}$ scheme)~\cite{Yang:2018nqn}. This difference will be further discussed below.

 With this in mind, the matrix elements of the split energy-momentum tensor are constrained by Lorentz symmetry

\bea
\label{T4}
\left<P|\bar T^{\mu\nu}|P\right>=&&2\bigg(P^\mu P^\nu-\frac 14g^{\mu\nu}M_N^2\bigg)\nonumber\\
\left<P|\hat T^{\mu\nu}|P\right>=&&\frac 12  g^{\mu\nu}M_N^2
\eea
The corresponding Hamiltonian in Minkowski signature, follows from the 00-component of (\ref{T1}-\ref{T3})
modulo BRST exact and gauge dependent contributions,

\bea
\label{T5}
H_G=\int d^3x\,\bar T^{00}_G&=&\int d^3x\,\frac 12(E^2+B^2)\nonumber\\
H^\prime_Q=\int d^3x\,\bar T^{00}_Q&=&\int d^3x\,\bigg(\frac 12 \overline\psi \gamma\cdot  i\overleftrightarrow D\psi+\frac 3{4}m\bar\psi \psi \bigg)\nonumber\\
H^\prime_A=\int d^3x\,\hat T^{00}_A&=&\int d^3x\,\frac 14 \bigg(\frac{\beta(g^2)}{4g^4}F^2+m\overline\psi\psi\approx 
-\frac {b}{32\pi^2}F^2+m\overline\psi\psi\bigg)
\eea
where the time  $t=0$  is subsumed. The mass term can be rearranged  so that (\ref{T5}) reads

\bea
\label{T5X}
H_G=\int d^3x\,\bar T^{00}_G&=&\int d^3x\,\frac 12(E^2+B^2)\nonumber\\
H_Q=\int d^3x\,\bar T^{00}_Q&=&\int d^3x\,\bigg(\frac 12 \overline\psi \gamma\cdot  i\overleftrightarrow D\psi\bigg)\nonumber\\
H_A=\int d^3x\,\hat T^{00}_A&=&\int d^3x\,\frac 14 \bigg(\frac{\beta(g^2)}{4g^4}F^2\approx 
-\frac {b}{32\pi^2}F^2\bigg)\nonumber\\
H_m=\int d^3x\,\bar T^{00}_G&=&\int d^3x\, m\overline\psi\psi
\eea
The nucleon mass budget is  then

\be
\label{T6}
M_N=\frac{\left<P|H_G+H_Q+H_A+H_m|P\right>}{\left<P|P\right>}\equiv M^N_G+M^N_Q+M^N_A+M^N_m
\ee
which shows that the combination

\be
M_{\rm inv}=M^N_G+M^N_Q+M^N_A
\ee
is  chirally symmetric and equal to the  invariant mass in (\ref{MASS}).

In Euclidean signature, whether on the lattice or using the QCD instanton vacuum, (\ref{T6}) can be evaluated by 
 trading $T^{00}\rightarrow T^{44}$ and $t=0\rightarrow i0$.
 In the dilute QCD instanton vacuum,
 the gluonic operator  in (\ref{T5}-\ref{T6}) is the sum of multi-instanton contributions of the form

\bea
\label{T7}
\bar T^{44}_G[A]=
\sum_{I=1}^{N_\pm }\bar T^{44}_G[A_I(\xi_I)]+
\sum_{I\neq J=1}^{N_{\pm }}\bar T^{44}_G[A_I(\xi_I), A_J(\xi_J)]+...=\sum_{I\neq  J=1}^{N_{\pm }}\bar T^{44}_G[A_I(\xi_I), A_J(\xi_J)]+...\nonumber\\
\eea
Since the first one-instanton contribution in (\ref{T7}) is composed of self-dual fields it vanishes. So we are left with only
the two and higher multi-instanton contributions. When averaged over a measure of independent instantons, the remaining
terms in (\ref{T7}) are suppressed by the diluteness factor  $\kappa\approx 0.1$.
 As a result, the contribution of $M^N_G$ is  parametrically small
 in comparison to $M^N_{Q}$ or $4M^N_A$, i.e.   $M^N_G/M^N_{Q}\approx \kappa\approx 0.1$.  The contributions $M^N_{Q,m}$ are  solely given in terms of the fermionic zero modes (modulo the instanton gauge fields  in the long derivative).

 With this in mind,  the breakdown in the mass budget (\ref{T6})  for the nucleon  yields the estimates

\bea
\label{T8}
\frac{M^N_Q}{M_N}&\approx &\frac 34 \frac 1{1+\kappa}\bigg(1-\frac{\sigma_{\pi N}}{M_N}\bigg)\approx  64\%\nonumber\\
\frac{M^N_G}{M_N}&\approx &\frac {3}4 \frac \kappa{1+\kappa}\bigg(1-\frac{\sigma_{\pi N}}{M_N}\bigg)\approx 7\% \nonumber\\
\frac{M^N_A}{M_N}&=&\frac {1}4\bigg(1-\frac{\sigma_{\pi N}}{M_N}\bigg) \approx 24\% \nonumber\\
\frac{M^N_m}{M_M}&=&\frac{\sigma_{\pi N}}{M_N}\approx 5\%
\eea
 with the empirical pion-nucleon sigma term (\ref{PIN}).
The anomalous contribution is scale
and scheme independent at one-loop order. The mass contribution is also renormalization group invariant.
(\ref{T8}) shows that in the QCD instanton vacuum, about 70\%  of the nucleon mass
stems from the valence quarks (hopping zero modes),  25\% from the gluon condensate or epoxy (displaced vacuum instanton field), 
and 7\% from emerging  valence gluons. The nucleon is composed mostly of quark constituents hopping and dragging the gluon epoxy.
The gluon epoxy in the nucleon is the quantum anomalous energy in the nucleon discussed recently in~\cite{Ji:2021pys}.

We note that the budgeting of the nucleon mass in
(\ref{T8}) differs from the one reported on the lattice in~\cite{Yang:2018nqn}, with a  noticeably  larger valence gluon fraction  in the lattice nucleon.
In our analysis, this can only be accomodated by  a  stronger instanton packing fraction of $\kappa\approx 0.5$ instead of 0.1, which is unlikely.  
(Note  that larger values of $\kappa$ that include close instanton-anti-instanton pairs, not responsible for the breaking of chiral symmetry, 
were reported when analyzing certain correlations at  zero cooling time~\cite{Athenodorou:2018jwu}). The
 harder renormalization scale $\mu=2\,{\rm GeV}$ used in the reported lattice results,  is the  likely  source of the valence and perturbative
gluon enhancement reported in the lattice nucleon. Quantum evolution will  enhance $M_G^N$ at the expense of $M_Q^N$, which in (\ref{T8})
would amount to effectively  dressing    $\kappa\approx 0.1\rightarrow 0.5$ at $\mu=2\,{\rm GeV}$. 

Finally, a similar mass decomposition holds for the pion at the same  soft renormalization scale of  $\bar\rho=0.6\,{\rm GeV}$, with the estimates

\bea
\label{T9}
\frac{M^\pi_Q}{m_\pi}&\approx &\frac 38 \frac 1{1+\kappa}\approx  34\%\nonumber\\
\frac{M^\pi_G}{m_\pi}&\approx &\frac {3}8 \frac \kappa{1+\kappa}\approx 3\% \nonumber\\
\frac{M^\pi_A}{m_\pi}&=&\frac {1}8\approx 13\% \nonumber\\
\frac{M^\pi_m}{m_\pi}&=&\frac 12\approx 50\%
\eea
About 85\% of the pion mass stems from the  valence  quarks (hopping zero modes), 13\%  from the gluon condensate or epoxy
(displaced vacuum instanton field), and 3\% from emerging valence gluons. Needless to say that all mass  contributions in the pion vanish 
smootly in the chiral limit. Again, quantum evolution will  enhance $M_G^\pi$ at the expense of $M_Q^\pi$, with effectively dressing 
 $\kappa\approx 0.1\rightarrow 0.5$ at $\mu=2\,{\rm GeV}$.


\section{Measuring the QCD vacuum compressibility}~\label{sec_compressibility}

While the present discussion has focused on some key aspects of the QCD vacuum and 
the hadronic mass sum rule, it is worth noting that the results (\ref{corr1}-\ref{corr2}) can be recast in the
following form 

\be
\label{VAC1}
\frac{\langle P\left|{F^2(0)}\right|P\rangle}{(4\pi (m_N-\sigma_{\pi N}/2))^2}\approx 
-\sigma_{F^2}\,
\ee
with the QCD vacuum compressibility

\be
\label{VAC2}
\sigma_{F^2}=\frac 1{32\pi^2}\int d^4x \frac{\langle F^2(x)\,F^2(0)\rangle_C}{\langle F^2(0)\rangle}
\ee
A measure of  the gluon condensate or epoxy  inside the proton (left hand-side) is a measure of the  QCD vacuum compressibility
$\sigma_{F^2}$ (right hand-side), modulo the pion-nucleon sigma  term which is small.   Since (\ref{VAC1}) is a nucleon connected matrix
element, it is natural that it probes the fluctuations of $F^2$. While  in the vacuum state the gluon condensate  is
positive,   (\ref{VAC2}) shows that it is negative in the nucleon state. The nucleon state carries less epoxy.

The cooled Yang-Mills vacuum in~Fig.~\ref{fig_VAC} is composed of interacting topological charges.
The  vacuum compressibility  $\sigma_{F^2}$ captures the squared variance of their interactions:
$\sigma_{F^2}=1$ for a  non-interacting gas phase, $\sigma_{F^2} <1$ for an interacting  liquid phase,  and $\sigma_{F^2}\ll 1$ for
a strongly interacting 
crystal phase.  QCD low-energy theorems suggest $\sigma_{F^2}\approx 4/b\approx 4/11$ (one-loop  and quenched)~\cite{Novikov:1981xi}, so the  QCD instanton
vacuum  appears to be a  dilute quantum topological liquid. A measure of $\sigma_{F^2}$ is a measure of a fundamental and universal parameter of the QCD vacuum.

\section{Conclusions}
\label{sec_conclusions}

The QCD
instanton vacuum is populated with topological tunneling configurations, with each  costing zero energy. The way a light quark can propagate
coherently through this maze of tunneling configurations is through its zero mode, scattering  and hopping from  an  instanton to an anti-instanton
and so on. The scattering through the instanton  flips chirality, an amazing effect  caused by  a non-perturbative vector interaction (a perturbative gluon 
interaction preserves chirality).
The hopping generates a very dense band in the virtual quark spectrum, reminiscent of the conduction band in conductors.
As a result, chiral symmetry is  spontaneously broken, a  chiral condensate is formed  and a running constituent quark mass emerges.

The QCD instanton vacuum breaks simultaneously conformal symmetry,  with a large and negative vacuum energy density, 
or  equivalently a large and positive gluon condensate (gluon epoxy).
A hadronic excitation in this vacuum, whether a quark, a meson or a baryon costs energy or mass. A useful and physical way to budget this mass is 
Ji$^\prime$s mass decomposition of the energy momentum tensor~\cite{Ji:1994av,Ji:1995sv}. In the QCD instanton vacuum, we find that the hadronic masses are  largely
due to the contribution  of the valence quarks as they hop and drag the gluon epoxy.

Finally,  a measure of the gluon condensate or epoxy in the nucleon,  is a measure  of  the compressibility of the QCD
instanton vacuum as a  topological liquid.  The diluteness of this liquid is central  in our  non-perturbative  understanding of the emergence of mass in QCD
using analytical methods. This gluonic content of the proton may be accessible through treshold electromagnetic production of heavy
quarkonia~\cite{Joosten:2018gyo}, and perhaps diffractive cluster production in hadron-hadron collisions~\cite{Shuryak:2021iqu}  at current and future facilities.

\vskip 1cm
{\bf Acknowledgements}

I thank Xiang-dong Ji, Zein-Eddine Meziani and Edward Shuryak for  discussion.
This work is supported by the Office of Science, U.S. Department of Energy under Contract No. DE-FG-88ER40388.

\appendix

\bibliography{twist3}

\end{document}